\documentclass{iau} 
\usepackage{graphicx}

\title[SO survey of massive cores] 
{SO survey of massive cores}

\author[Igor Zinchenko \& Christian Henkel]   
{Igor Zinchenko$^1$
 \and Christian Henkel$^2$}

\affiliation{$^1$Institute of Applied Physics of the Russian Academy of Sciences,\\
	   46 Uljanov~str., 603950 Nizhny Novgorod, Russia \\ email: {\tt zin@appl.sci-nnov.ru} \\[\affilskip]
$^2$Max-Planck-Institut f\"ur Radioastronomie, \\ Auf dem H\"ugel 69,
D-53121 Bonn, Germany \\email: {\tt chenkel@mpifr-bonn.mpg.de}}

\pubyear{2017}
\volume{332}  
\jname{Astrochemistry VII –-- Through the Cosmos from Galaxies to Planets}
\editors{Maria Cunningham, Tom Millar \& Yuri Aikawa, eds.}
\begin{document}

\maketitle

\begin{abstract}
We present the results of a survey of several tens of dense high mass star forming (HMSF) cores in three transitions of the SO molecule at 30 and 100 GHz with the 100-m Effelsberg and 20-m Onsala radio telescopes. The physical parameters of the cores are estimated from the line ratios and column densities. Relative abundances are derived as well.

\end{abstract}

\section{Introduction}
The chemistry of sulfur-bearing molecules is still poorly explored, especially in high mass star forming regions. These molecules can be important tracers of various phenomena in star forming regions, including outflows. Relative abundances of sulfur-bearing molecules can serve as chemical clocks \cite[(e.g. Viti et al. 2004)]{Viti04} and as an indicator of the initial C/O ratio \cite[(e.g. Bergin et al. 1997)]{Bergin97}. However, there are some discrepancies between models and observations \cite[(e.g. Herpin et al. 2009)]{Herpin09}

\section{Observations and data analysis}
We have observed 3 transitions of the SO molecule:
$J_N = 1_0 - 0_1$ at 30.0016~GHz (with the 100-m Effelsberg telescope), 
$J_N = 3_2 - 2_1$ at 99.2999~GHz and simultaneously $J_N = 4_5 - 4_4$ at 100.0296~GHz (with the 20-m Onsala telescope).
The beam widths of the Effelsberg and Onsala antennas at these wavelengths are similar ($\sim 30^{\prime\prime}$ and $\sim 40^{\prime\prime}$, respectively).

We analyze the line ratios with RADEX \cite[(Van der Tak et al. 2007)]{RADEX}. The line ratios were modeled in wide ranges of density and temperature. Estimates of the density and temperature from the line ratios are used for estimates of the SO column densities.
SO abundances are derived by comparison with the C$^{18}$O column densities, found from our Onsala observations \cite[(Zinchenko et al. 2000)]{Zin00}. The relative C$^{18}$O abundance is assumed to be $X(\mathrm{C^{18}O}) = 1.7\times 10^{-7}$ \cite[(Frerking et al. 1982)]{Frerking82}.

\section{Results}
\begin{figure}[htb]
\centering
\includegraphics[width=\textwidth]{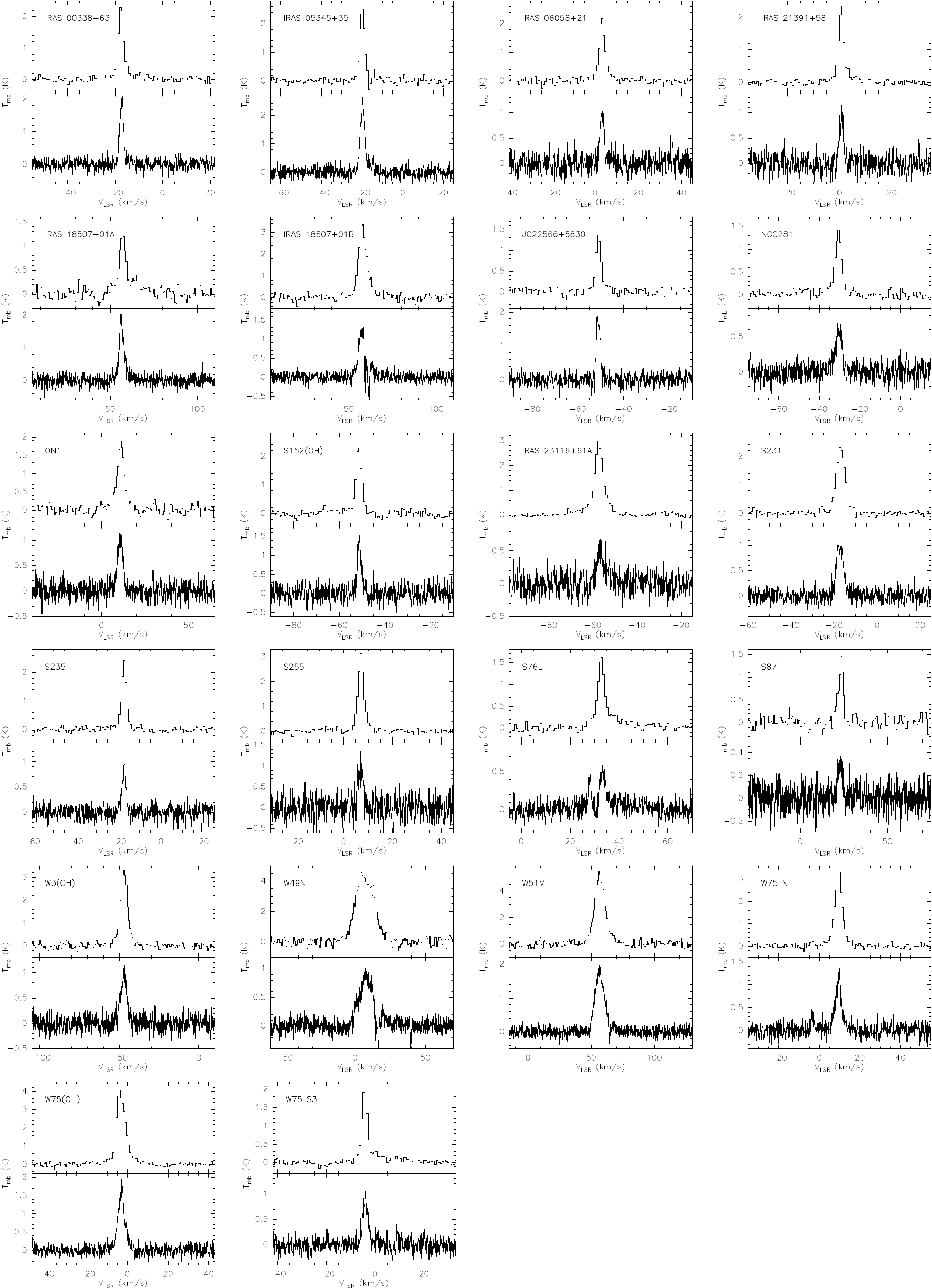} 
\caption{The $J_N = 3_2 - 2_1$ (upper panels) and $J_N = 1_0 - 0_1$ (lower panels) spectra for the 22 sources observed both in Effelsberg and in Onsala.}
\label{fig:summary}
\end{figure}

At Onsala we have observed 35 sources. The $J_N = 3_2 - 2_1$ transition has been detected in all of them. The $J_N = 4_5 - 4_4$ transition has been detected in 3 objects: W3 (OH), W51 M and ON2 N.
In Effelsberg we have observed 22 sources from the same sample. The SO line has also been detected in all of them. The $J_N = 3_2 - 2_1$ and $J_N = 1_0 - 0_1$ spectra for these sources are presented in Fig.~\ref{fig:summary}.

\begin{figure}
\centering
\includegraphics[width=0.9\textwidth]{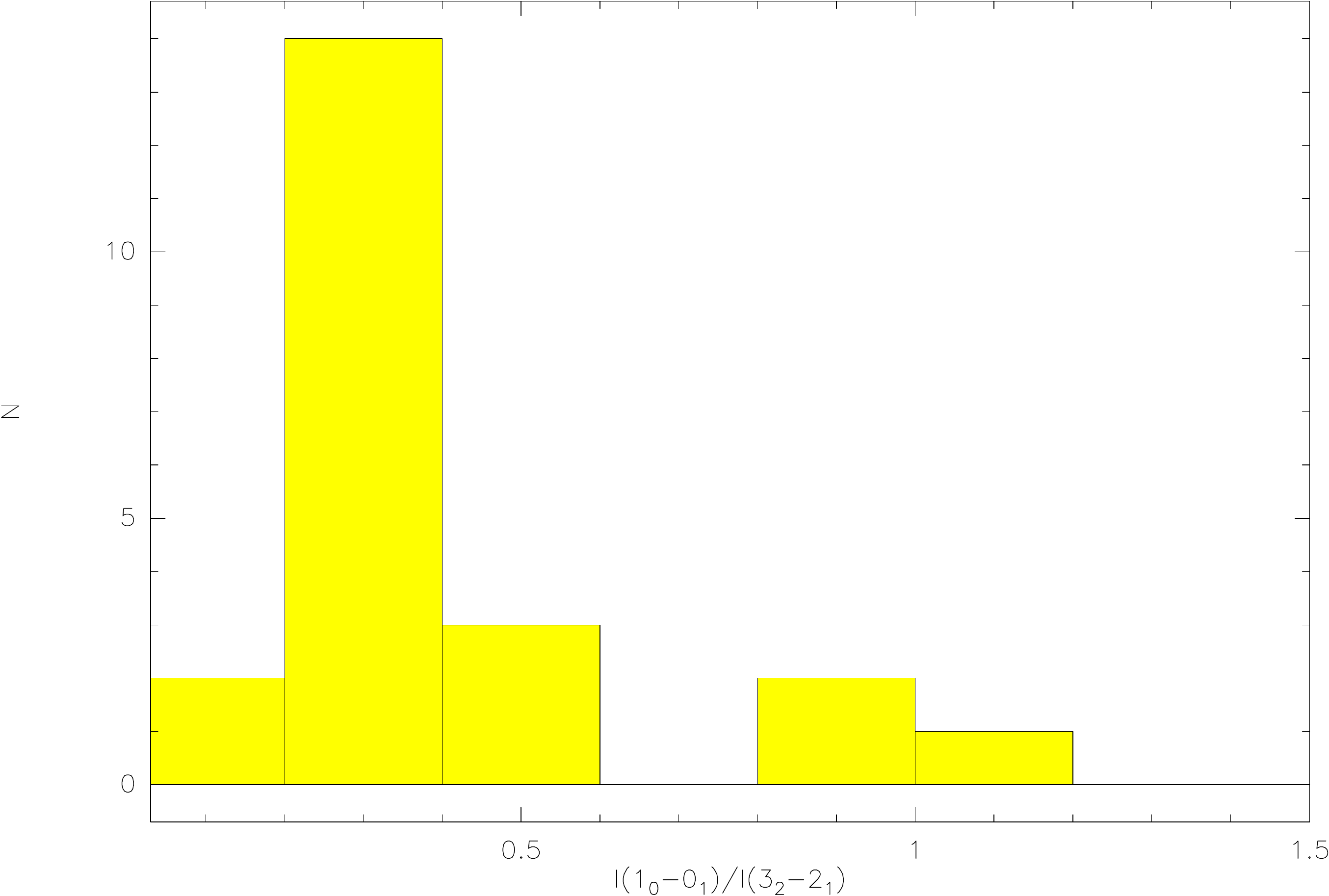} 
\caption{The histogram of the $(1_0 - 0_1)/(3_2 - 2_1)$ line intensity ratio.}
\label{fig:r1-hist}
\end{figure}

The $(1_0 - 0_1)/(3_2 - 2_1)$ intensity ratio lies in the range from $\sim 0.15$ to $\sim 2$ (Fig.~\ref{fig:r1-hist}). This corresponds to densities from $\sim 10^6$~cm$^{-3}$ to $\sim 5\times 10^3$~cm$^{-3}$. Most of the sources have a ratio of $ \sim 0.3 $, which corresponds to a density of $\sim 10^5$~cm$^{-3}$. 
The $(3_2 - 2_1)/(4_5 - 4_4)$ intensity ratio is $\sim 20$ for the detected sources. This detection indicates high temperature ($> 50$~K) and density ($> 10^6$~cm$^{-3}$).

\begin{figure}
\centering
\includegraphics[width=0.9\textwidth]{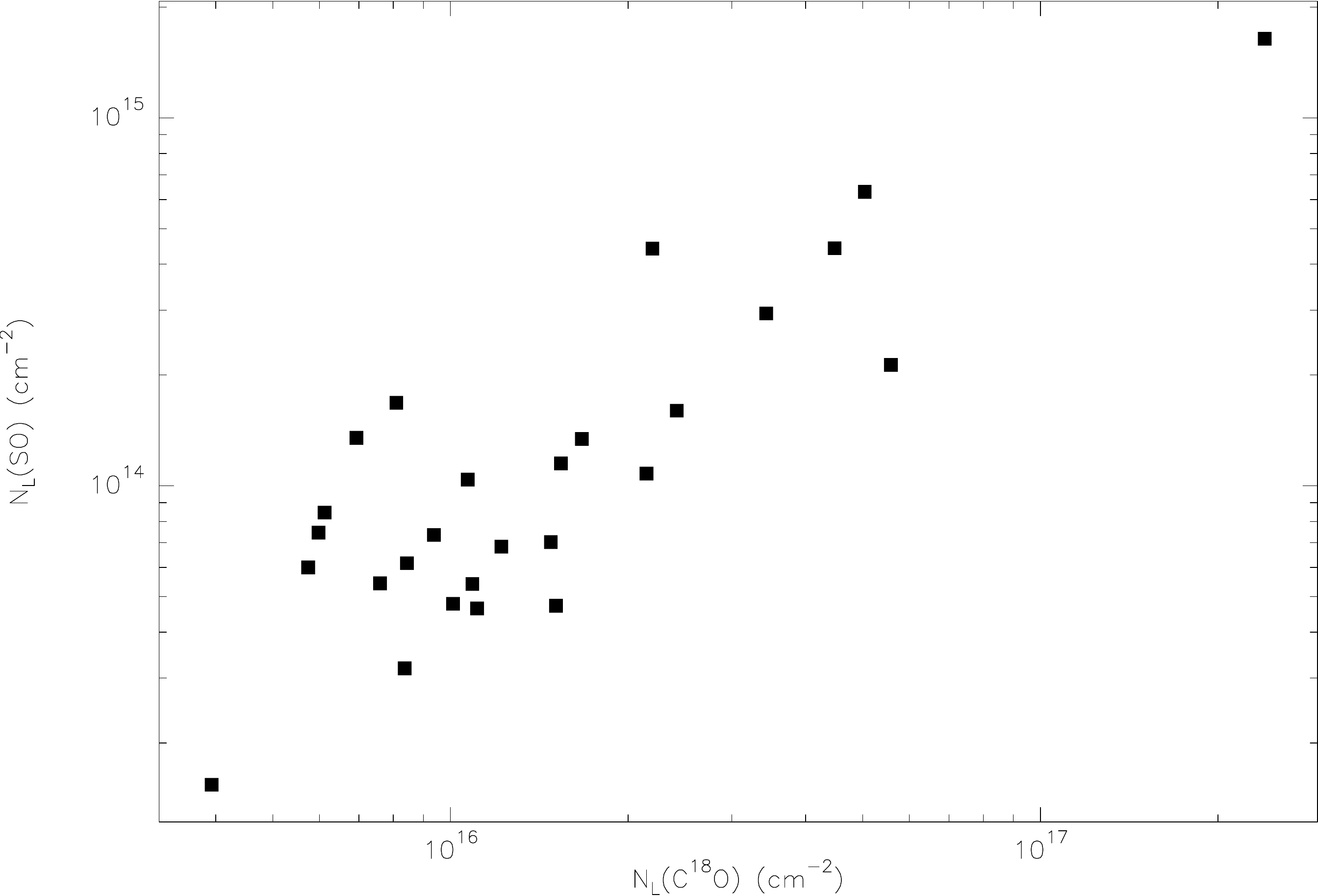} 
\caption{The SO column density in dependence on the C$^{18}$O column density.}
\label{fig:so-c18o}
\end{figure}

There is a good correlation between the SO and C$^{18}$O column densities (Fig.~\ref{fig:so-c18o}).

\begin{figure}
\centering
\includegraphics[width=0.9\textwidth]{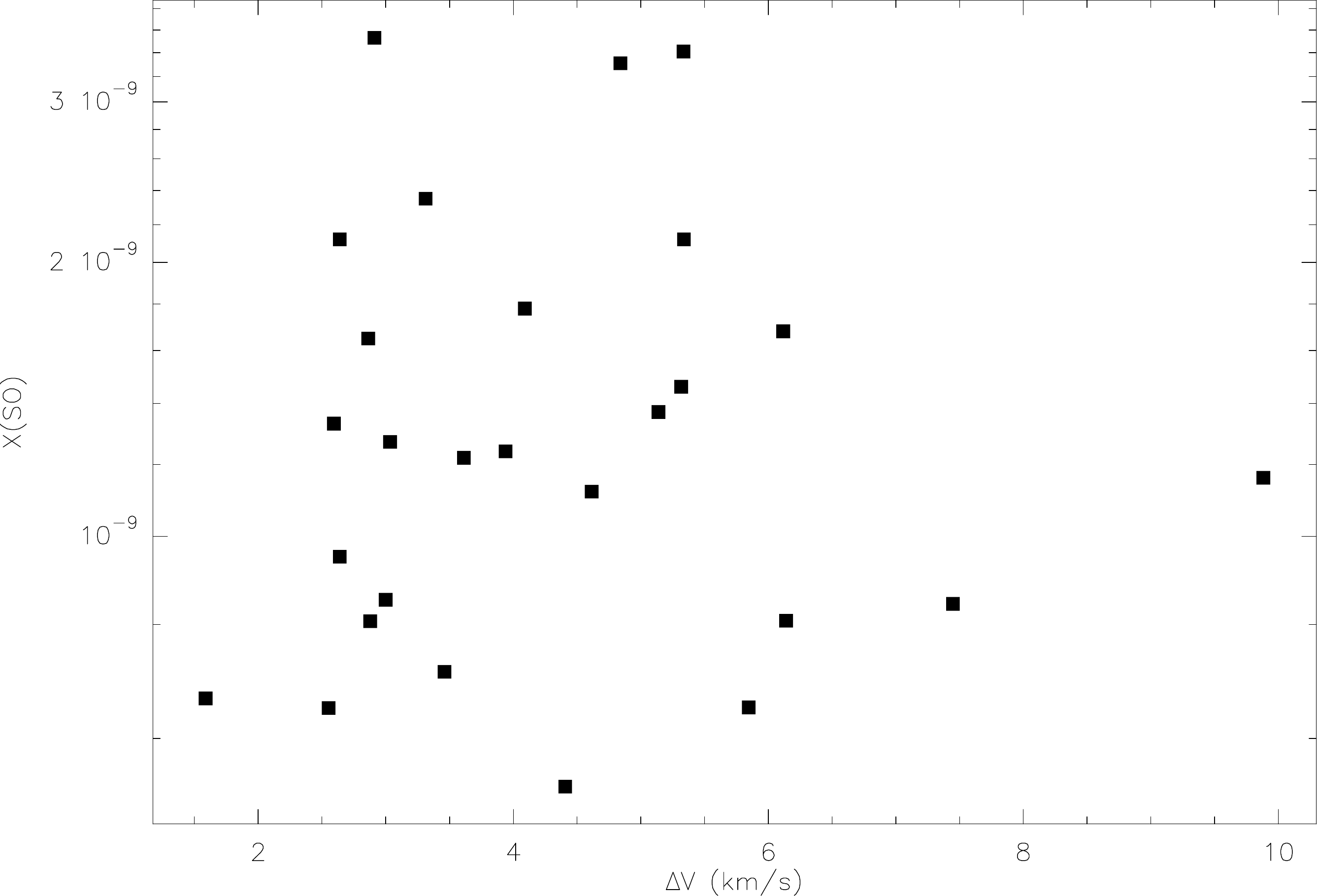} 
\caption{The SO relative abundance in dependence on the SO line width.}
\label{fig:x_so-dv}
\end{figure}

The SO relative abundance varies in the range $X(\mathrm{SO}) \sim (0.5 - 4)\times 10^{-9}$ with a median value of $\sim 1.3\times 10^{-9}$. These values are rather close to the results of other recent studies of similar objects \cite[(e.g. Li et al. 2015)]{Li15}. We see no correlation between the SO abundance and the line width (Fig.~\ref{fig:x_so-dv}).

\section{Conclusions}
We have surveyed 35 HMSF cores in the SO lines at 3 mm and 1 cm wavelengths. 22 objects have been observed (and detected) at both wavelengths.
The line intensity ratios indicate a typical density of $\sim 10^5$~cm$^{-3}$. For several objects detected in the $J_N = 4_5 - 4_4$ line the density should be $> 10^6$~cm$^{-3}$ and temperature $> 50$~K.
There is a good correlation between the SO and C$^{18}$O column densities. 
The SO relative abundance varies in the range $X(\mathrm{SO}) \sim (0.5 - 4)\times 10^{-9}$ with a median value of $\sim 1.3\times 10^{-9}$. There is no correlation between the $X(\mathrm{SO})$ and line width.

This research was supported by the Russian Foundation for Basic Research (recently by the grant No. 15-02-06098) and by the IAP RAS state program 0035-2014-0030.

\end{document}